\documentclass[pra,showpacs,showkeys]{revtex4} \usepackage{latexsym}
\usepackage{amsfonts} \usepackage{amsmath,revsymb} 

\newcommand{\kt}[1]{\ensuremath{|#1\rangle}}
\newcommand{\br}[1]{\ensuremath {\langle #1|}}
\newcommand{\bk}[2]{\ensuremath {\langle #1|#2 \rangle}}
\newcommand{\kb}[2]{\ensuremath {| #1 \rangle\langle #2|}}
\newcommand{\Tr}{\mathrm{Tr}}
\newcommand{\HS}{\mathcal{H}}
\newcommand{\vp}{\mathbf{p}}
\newcommand{\vP}{\hat{\mathbf{P}}}
\newcommand{\vz}{\mathbf{z}}
\newcommand{\vZ}{\hat{\mathbf{Z}}}
\newcommand{\vk}{\mathbf{k}}

\begin{document}

\title{Entanglement Generation in the Scattering of One-Dimensional Particles}

\author{N.L.~Harshman and  G.~Hutton}
\affiliation{Department of Computer Science, Audio Technology, and Physics\\
4400 Massachusetts Ave., NW \\ American University\\ Washington, DC 20016-8058}

\begin{abstract}
This article provides a convenient framework for quantitative evaluation of the entanglement generated when two structureless, distinguishable particles scatter non-relativistically in one dimension. It explores how three factors determine the amount of entanglement generated: the momentum distributions of the incoming particles, their masses, and the interaction potential.  Two important scales emerge, one set by the kinematics and one set by the dynamics.  This method also provides two approximate analytic formulas useful for numerical evaluation of entanglement and reveals an interesting connection between purity, linear coordinate transformations, and momentum uncertainties.
\end{abstract}

\maketitle

\section{Introduction}

Two structureless, distinguishable particles scatter non-relativistically in one dimension.  Specifically, the two particles are prepared independently, interact via a potential with finite range, and separate again.  When particles with no initial classical or quantum correlations scatter, they generally emerge from the interaction region entangled. In this context we ask the question: how much entanglement between the particles is generated and on what does it depend?  The topic of entanglement in particle scattering has intrinsic interest: scattering is a fundamental dynamical process and entanglement is the fundamental resource for quantum information processes.  Despite the potentially complicated dependence of entanglement generation on the momentum distributions of the incoming particles, their masses, and the interaction potential, this article identifies some general features of entanglement generated in scattering systems.

Entanglement generation in non-relativistic scattering of structureless, distinguishable particles has been considered previously~\cite{schulman98,mack02,schulman04,law04,tal05,wang05, wang06, janzig06,frey07}.  Most previous treatments consider interactions of a single species~\cite{mack02,law04,tal05,wang05,wang06,frey07}.  Some consider scattering by particles confined in a harmonic well interacting by delta-function potentials~\cite{mack02,wang05,frey07} and were directly inspired by experiments with ultracold atoms in traps.  Analytic results for the energy levels of such systems~\cite{busch98,cirone01} can be used as a discrete basis for calculating entanglement.  Others have considered Gaussian wave functions of unbounded particles interacting via specific limited-range potential models, including hard core~\cite{schulman98,schulman04,law04,janzig06}, ``cavity'' (hard core plus delta-function)~\cite{law04}, and double delta-function~\cite{tal05}.  
Each of these have considered different aspects and have particular strengths.  This research seeks to unify some of the different observations of these papers and provide a useful framework for analyzing general properties of entanglement generation in scattering.

Scattering also supplies a rich context for exploring techniques of quantum information theory with continuous variable systems.   A bipartite continuous variable pure state will have a wave function $f(x_1,x_2)$.  The state is unentangled if the wave function is separable, and then it can be written as $f(x_1,x_2)=g(x_1)h(x_2)$.  An entanglement measure will quantify the non-separability.  Some inseparable wave functions can be written as
\begin{equation}\label{mode}
f(x_1,x_2)=\sum_{i=1}^N f_i(x_1,x_2)
\end{equation}
where each of the $f_i$ have disjoint domains of support and are therefore orthogonal.  Each $f_i$ can be thought of as the wave function of a ``mode'' of the system, and entanglement due to the superposition of modes can be called ``mode-like''~\cite{harshqic07}. Either the modes $f_i$ are   separable, and then the only entanglement is mode-like, or they are themselves irreducibly inseparable.  Use the purity of the one-particle reduced density matrix $p_{12}$ defined below, and the purity of wave function like (\ref{mode}) is
\begin{equation}\label{modepur}
p_{12}(f)=\sum_{i=1}^N p_{12}(f_i).
\end{equation}
In other words, the purity of a state that can be reduced into orthogonal modes is the sum of the purities of those modes.

Using these ideas, several different analytic expressions for the purity will be given for different approximate or incomplete-knowledge situations that arise in particle scattering.  Scattering can split the wave function into modes, such as a transmission mode and a reflection mode, or it can distort the wave function shape.  Either effect can render it inseparable.  In resonance scattering~\cite{law04,tal05}, the rapid variation of the scattering amplitude with relative momentum leads to such distortion, but even when the amplitudes are constant, the wave function can be rendered inseparable by the distortion caused by reflection.  Entanglement that originates from kinematic effects, like entanglement due to reflection, can be distinguished from dynamic effects using the methods described here.

As an example, consider hard core repulsion in one-dimension.  In \cite{law04}, the case of two particles with the same mass is studied.  The amount of entanglement is measured by the inverse of the purity of the one-particle reduced density matrix.  Its variation with time is numerically calculated, and first it grows from no entanglement to a maximum value as the particles approach and interact.  Then it recedes back to no entanglement.  In contrast, as first shown in \cite{schulman98} and also discussed in \cite{schulman04,janzig06}, if the particles have \emph{different} masses, then the outgoing states will remain entangled after scattering, and possibly very entangled, unless
\begin{equation}\label{sec}
\frac{m_1}{\sigma^2_1}=\frac{m_2}{\sigma^2_2},
\end{equation}
where $\sigma_i = \Delta p_i$ is the momentum uncertainty\footnote{Note that in \cite{schulman98,schulman04,janzig06}, position uncertainties are used instead of the momentum uncertainties and so their $\sigma$ is the inverse of our $\sigma$.}. In one-dimensional scattering  hard core repulsion means only reflection can occur, and this leads to the following interpretation:  During the time interval the scattering actually occurs, even with equal masses, the two-particle wave function is distorted by interference between the parts of the wave function that are still incoming and the reflected parts.  For the equal mass case, after a long time the final wave function will return to an undistorted separable state.  However, if the masses are different, the process of reflection will engender correlations between the particles that remain after the particles leave the scattering region.  Reflection means reversing the internal momentum, but different parts of the single-particle momenta distributions will be affected in different ways by this transformation and so the wave function will be distorted.  More details are given below.

While the approach of this article uses the S-matrix, thus excluding exploration of the time dependence of the entanglement, it does allow the separation of kinematic and dynamic entanglement effects by exploiting symmetries and conservation laws.  Another critical feature of our approach to entanglement in one-dimensional scattering is the application of the mathematics of tensor product structures (TPSs).  The notion of entanglement depends on how a system is factored into subsystems.  Most treatments of entanglement in scattering are concerned exclusively with interparticle entanglement, in other words entanglement with respect to the TPS
\begin{equation}\label{iTPS}
\HS=\HS_1\otimes\HS_2,
\end{equation}
where $\HS_i$ is the Hilbert space of a single particle.  Other TPSs are useful for scattering entanglement. For example, one can define a TPS which factors the into the internal and external degrees of freedom (see \cite{harshprl07,harshosid07}).  In \cite{harshprl07} it was proved that for non-relativistic scattering, entanglement with respect to the internal-external TPS is unaffected by scattering because Galilean invariance interactions will factor into unitary operators operating on each product space.  Generally, a change of observables leads to a new notion of entanglement, and this can occasionally be exploited to achieve useful analytic and approximate results.  In particular, the Schulman condition (\ref{sec}) can be derived as a special case of the purity of Gaussian wave functions under linear transformations of the observables, and the entangling distortion of reflection can be analytically calculated in the same framework.

The outline of this paper is as follows:  First, we describe the how to quantify entanglement for two particles in one dimension using the purity of the reduced density matrix, and we show how coordinate transformations lead to new TPSs and new definitions of entanglement.  Then we  describe the scattering dynamics and show how the boundary conditions of scattering imply that the total entanglement of the scattered state is partially mode-like, in the sense that it is the sum of the the entanglement of the transmitted and reflected states.  Next we consider several different analytic approximations to the purity and the conditions under which they apply.  In the process, three different aspects of dynamical entanglement will be thrown into distinction, entanglement due to superposition of transmission and reflection, entanglement due to wave packet distortion under reflection, and entanglement due to wave packet distortion due to convolution with a rapidly-varying scattering amplitude.  At the end, we will comment on possible applications and indicate directions for further research.

\section{Quantifying entanglement of two one-dimensional particles}

For a system of two spinless, structureless, non-relativistic particles in one dimension, a complete set of commuting observables (CSCO) is  the momentum operators for each particle $\{\hat{P}_1,\hat{P}_2\}$.  
The generalized eigenvectors (eigenkets) of these operators are $\kt{p_1,p_2}$ and they have delta-function ``normalization''
\begin{equation}
\bk{p_1,p_2}{p_1',p_2'}=\delta(p_1-p_1')\delta(p_2-p_2').
\end{equation}
A general state of the two-particle system $\kt{\Phi}\in\HS$ can be realized by its two-particle momentum wave function $\phi(p_1,p_2)$:
\begin{equation}
\kt{\phi} = \int dp_1dp_2 \phi(p_1, p_2) \kt{p_1,p_2}.
\end{equation} 
The CSCO used above implies a tensor product structure, i.e.\ a way of partitioning the total Hilbert space
\begin{equation}\label{eq:tp12}
\HS = \HS_1 \otimes \HS_2 \sim L^2(\mathbb{R})\otimes L^2(\mathbb{R}) = L^2(\mathbb{R}^2),
\end{equation}
where $L^2(\mathbb{R})$ is the space of Lebesgue square-integrable functions~\footnote{Actually, as is normal in quantum mechanics, we will only consider elements of a dense subset $\Phi\subset \HS$ whose elements are realized by wave functions that are well-behaved, i.e. smooth, rapidly-decreasing.  An example is the Schwartz space.  This restrictions allows the use of the Dirac basis vector expansion and the use of Riemann integrals instead of Lebesgue integrals in calculations.}.
In turn, this tensor product structure (TPS) implies a notion of entanglement: interparticle entanglement.  

To quantify the interparticle entanglement, we use the purity of the one-particle reduced density matrix $p_{12}$, which for the sake of brevity we refer to as the interparticle purity:
\begin{equation}\label{eq:prep12}
p_{12}(\phi) = \Tr_1(\hat{\rho}_1^2) = \Tr_2(\hat{\rho}_2^2)
\end{equation}
The one-particle reduced density matrix is
\begin{equation}\label{eq:redmat}
\hat{\rho}_1 = \Tr_2(\hat{\rho})=\int dp_1dp_1'dp_2 \phi(p_1,p_2)\phi^*(p_1',p_2) \kb{p_1}{p_1'}.
\end{equation}
Inserting (\ref{eq:redmat}) into (\ref{eq:prep12}) gives a simple, symmetric expression for the interparticle purity:
\begin{equation}\label{eq:mompur}
p_{12}(\phi) = \int dp_1dp_2dp'_1dp'_2 \phi(p_1, p_2)\phi^*(p'_1, p_2)\phi(p'_1, p'_2)\phi^*(p_1, p'_2).
\end{equation}
Assuming that the pure state $\kt{\phi}$ is normalized, the purity will be in the set $p_{12}\in (0,1]$.  For a separable state $\phi_s(p_1, p_2)=\phi_1(p_1)\phi_2(p_2)$, one calculates $p_{12}=1$:
\begin{eqnarray}
p_{12}(\phi_s) &=& \int dp_1dp_2dp'_1dp'_2 \phi_1(p_1)\phi_2(p_2)\phi_1^*(p'_1)\phi_2^*(p_2)\phi_1(p'_1)\phi_2(p'_2)\phi_1^*(p_1)\phi_2^*(p'_2)\nonumber\\
&=& \left(\int dp_1 |\phi_1(p_1)|^2\right)^2 \left(\int dp_2 |\phi_2(p_2)|^2\right)^2.
\end{eqnarray}
One could imagine calculating the entropy of entanglement $E_{12}=-\mathrm{Tr}\hat{\rho}_i\log\hat{\rho}_i$ for a pure continuous-variable state, but diagonalizing the one-particle reduced density matrix is less amenable to producing analytic results.  To diagonalize, either one  calculates the eigenvalues of the continuous reduced density matrix, which is typically a hard problem, or one discretizes the momentum spectrum in some way, which could obfuscate some of the very effects we are trying to uncover.  

Other CSCOs are compatible with interparticle entanglement.  For example, the CSCO of the particle position operators $\{\hat{X}_1,\hat{X}_2\}$, leads to the same TPS (\ref{eq:tp12}) as the CSCO $\{\hat{P}_1,\hat{P}_2\}$.  The CSCOs and (therefore the wave functions) are related by Fourier transformations.  These are local unitary operators with respect to the interparticle TPS (\ref{eq:tp12}).  The interparticle entanglement of a position wave function $\tilde{\phi}(x_1,x_2)$ can calculated by Fourier transforming the position wave function to the momentum variables and the evaluating the purity $p_{12}$, or equivalently by evaluating
\begin{equation}\label{eq:pospur}
p_{12}(\phi) = \int dx_1dx_2dx'_1dx'_2 \tilde{\phi}(x_1, x_2)\tilde{\phi}^*(x'_1, x_2)\tilde{\phi}(x'_1, x'_2)\tilde{\phi}^*(x_1, x'_2).
\end{equation}

Similarly, global translations in position space and global boosts in momentum space are local unitary operators with respect to the interparticle TPS.  These coordinate transformations will be represented on (\ref{eq:tp12}) by non-entangling operators because the generators of these transformations factor into a direct sum of local operators.  For example, a global translation in space by $a$ is generated by the total momentum operator $\hat{P} = \hat{P}_1\otimes \hat{\mathbb{I}}_1 + \hat{\mathbb{I}}_2\otimes \hat{P}_2$.
When this is exponentiated, the translation operator factors into local unitaries: $\exp(ia\hat{P}_1)\otimes\exp(ia\hat{P}_2)$~\footnote{A local translation like $\exp(ia_1\hat{P}_1)\otimes\exp(ia_2\hat{P}_2)$ is also non-entangling, but this operator does not commute with the interaction Hamiltonian and  it is not a generator of the Galilean symmetry group for the whole system.}.  Similarly, one could boost the system without generating interparticle entanglement, including a global boost to the frame where the expectation value for the total momentum $\hat{P}$ is zero, i.e.\ the center-of-mass (COM) reference frame.  However, the interaction Hamiltonian does not separate into a sum of local operators on the interparticle TPS (\ref{eq:tp12}) since the interaction potential depends on the relative positions.  As a result, time translation, and therefore the scattering operator, are not local operators and can change the amount of interparticle entanglement.

There are other CSCOs which are useful for understanding the scattering of two particles in one dimension, but which are not compatible with the TPS (\ref{eq:tp12}).  For example, the COM and relative position coordinates 
\begin{subequations}\label{comtr}
\begin{eqnarray}
x_c &=& \mu_1 x_1 + \mu_2 x_2\\
x_r &=& x_2 - x_1\nonumber
\end{eqnarray}
and conjugate momentum coordinates
\begin{eqnarray}\label{spmtoiem}
p &=& p_1 + p_2\\
q &=& \mu_2 p_1 - \mu_1 p_2,\nonumber
\end{eqnarray}
\end{subequations}
where $\mu_i = m_i/(m_1 + m_2)$, are the natural coordinates for most scattering calculations.  The COM-relative momentum CSCO
\begin{equation}
\{\hat{P}= \hat{P}_1\otimes \hat{\mathbb{I}}_1 + \hat{\mathbb{I}}_2\otimes \hat{P}_2, \hat{Q}=\mu_2 \hat{P}_1\otimes \mathbb{I}_2 - \mu_1 \mathbb{I}_1\otimes \hat{P}_2\}
\end{equation}
defines the internal-external (IE) TPS for one-dimensional particles
\begin{equation}\label{eq:tpie}
\HS = \HS_c \otimes \HS_r \sim L^2(\mathbb{R})\otimes L^2(\mathbb{R}) = L^2(\mathbb{R}^2),
\end{equation}
and the COM-relative position observables $\{\hat{X}_c,\hat{X}_r\}$ define an equivalent TPS.

Note that the total Hilbert space in (\ref{eq:tp12}) and (\ref{eq:tpie}) is the same; only the partition into subsystems is different.  The subsystem partitions (\ref{eq:tp12}) and (\ref{eq:tpie}) are isomorphic, but they lead to inequivalent definitions of entanglement.  The interparticle wave function can be transformed into the IE wave function by
\begin{eqnarray}\label{momtoie}
\check{\phi}(p, q) &=& \int dp_1dp_2\phi(p_1, p_2) \bk{p,q}{p_1,p_2}\nonumber\\
&=& \int dp_1dp_2  \phi(p_1, p_2)\delta(p_1 - \mu_1 p - q)\delta(p_2 - \mu_2 p + q)\nonumber\\
&=& \phi(\mu_1 p + q, \mu_2 p - q).
\end{eqnarray}
As can be seen from the above expression, a pure state $\kt{\phi}$ that is separable with respect to the single-particle TPS (\ref{eq:tp12}) may not be separable with respect to the IE TPS (\ref{eq:tpie}) and vice versa.  

The IE TPS is useful because the Hamiltonian for all Galilean invariant interactions will have the form
\begin{eqnarray}
\hat{H} &=& \hat{P}^2/(2M) + \hat{Q}^2/(2m)+V(\hat{X}_r)\nonumber\\
&=& \hat{H}_c\otimes \mathbb{I}_1 + \mathbb{I}_1\otimes\hat{H}_r
\end{eqnarray}
where $M = m_1 + m_2$ and $m=m_1m_2/M = \mu_1\mu_2M$.  This separation of variables is exactly why this coordinate system is used to solve the Schr\"odinger equation, but it also can be seen as a statement about the dynamical entanglement of scattering.  Since the Hamiltonian is a sum of local operators with respect to the IE TPS (\ref{eq:tpie}), the entanglement between the relative and COM degrees of freedom (or between the internal and the external degrees of freedom) is a dynamical invariant for any Galilean invariant interaction.  This is a result that also holds for any Galilean invariant interaction in three dimensions and for particles that have internal structure such as spin, although the definition of the internal and external CSCO changes~\cite{harshprl07,harshosid07}.

The transformation of the momentum variables (\ref{spmtoiem}) is just one example of a real linear transformation of the CSCO.  Using the notation $\vp=(p_1, p_2)$ to denote a vector in $(p_1,p_2)$-space, we consider how to calculate the entanglement of a given wave function $\phi(p_1,p_2)=\phi(\vp)$ with respect to new variables $\vz=\mathrm{T}\vp$, where $\mathrm{T}$ is a real, $2\times 2$ matrix with $\det\mathrm{T}=\pm 1$ to preserve normalization.  These variables are the eigenvalues of a CSCO $\{\hat{Z}_1,\hat{Z}_2\}$ given by the transformed momentum operators $\vZ=\mathrm{T}\vP$.  This CSCO in turn implies a new TPS
\begin{equation}\label{zTPS}
\HS=\HS_{z_1}\otimes\HS_{z_2}\sim L^2(\mathbb{R})\otimes L^2(\mathbb{R}) = L^2(\mathbb{R}^2).
\end{equation}
and therefore to a new notion of entanglement is measured by a transformed purity.  The purity with respect to the new TPS is
\begin{equation}\label{eq:zpur1}
p_{\mathrm{T}12}(\phi) = \int dz_1dz_2dz'_1dz'_2 \check{\phi}(z_1, z_2)\check{\phi}^*(z'_1, z_2)\check{\phi}(z'_1, z'_2)\check{\phi}^*(z_1, z'_2),
\end{equation}
where $\check{\phi}(z_1,z_2) = \phi((T^{-1}\vz)_1,(T^{-1}\vz)_2)$ and $\vz=(z_1,z_2)$.

As an example, consider a state that is the product of two Gaussian distributions of momentum:
\begin{equation}\label{gaussmom}
\phi_G({\bf p}) =  N_1N_2 e^{ip_1 a_1}e^{-\frac{(p_1 - k_1)^2}{4\sigma^2_1}} e^{ip_2 a_2}e^{-\frac{(p_2 - k_2)^2}{4\sigma^2_2}},
\end{equation}
where $N_i = (2\pi\sigma^2_i)^{-1/4}$, $k_i$ are the central momenta, $a_i$ are the central positions, and $\sigma_i$ are the momentum uncertainties for each particle's Gaussian.  Because it was constructed as a product of single particle wave functions, the interparticle purity of this state  $p_{12}=1$.  
However, the transformed purity (\ref{eq:zpur1}) of $\kt{\phi_G}$ for a transformation matrix
\begin{equation}
{\rm T} = \left(\begin{array}{cc} r & s\\ t & u\end{array} \right),
\end{equation}
where $|ru-st|=1$, is
\begin{equation}\label{pur_trans1}
p_{\mathrm{T}12}(\phi_G) = \frac{\sigma_1 \sigma_2}{\sqrt{(r^2\sigma_1^2 + s^2\sigma_2^2)(t^2\sigma_1^2 + u^2\sigma_2^2)}}.
\end{equation}
See the Appendix for the derivation.  The quantity (\ref{pur_trans1}) does not disappear even when momentum the variances $\sigma_i$ are small.  The scale for this kind of entanglement is set by the ratio of the momentum uncertainties, as can be seen more clearly by defining $c=\sigma_2/\sigma_1$:
\begin{equation}\label{pur_trans2}
p_{\mathrm{T}12}(\phi_G) = \frac{c}{\sqrt{(r^2 + s^2c^2)(t^2 + u^2c^2)}}.
\end{equation}

Specifying to the matrix that transforms from the particle momentum coordinates to the IE CSCO,
\begin{equation}
{\rm T}_{cm} = \left(\begin{array}{cc} 1 & 1\\ \mu_2 & -\mu_1\end{array} \right),
\end{equation}
 we find the IE entanglement is
\begin{equation}\label{pur_com}
p_{IE}(\phi_G) = \frac{\sigma_1 \sigma_2}{\sqrt{(\sigma_1^2 + \sigma_2^2)(\mu_2^2\sigma_1^2 + \mu_1^2\sigma_2^2)}}.
\end{equation}
Inspecting this function, one finds that a separable Gaussian state will have entanglement with respect to the IE TPS unless the condition
\begin{equation}
\frac{\mu_1}{\sigma_1^2} = \frac{\mu_2}{\sigma_2^2}
\end{equation}
is satisfied.  This is equivalent to the Schulman condition (\ref{sec}) which was found by explicitly considering the transformed wave function~\cite{schulman98}
\begin{eqnarray}
\check{\phi}_G(p,q)&=&N_1N_2 e^{i(\mu_1 a_1+\mu_2 a_2)p}e^{-\left(\frac{\mu_1^2}{4\sigma_1^2}+\frac{\mu_2^2}{4\sigma_2^2}\right)(p-k_1-k_2)^2}\nonumber\\
&&\times e^{-\left(\frac{1}{4\sigma_1^2}+\frac{1}{4\sigma_2^2}\right)(q-\mu_2 k_1 + \mu_1 k_2)^2}e^{-\left(\frac{\mu_1}{4\sigma_1^2}-\frac{\mu_2}{4\sigma_2^2}\right)(p-k_1-k_2)(q-\mu_2 k_1 + \mu_1 k_2)}
\end{eqnarray}
and looking for the separability conditions.  The relevance of this fact will be seen when the kinematics and dynamics of scattering are discussed below.

\section{Dynamical entanglement through scattering} 

The in-state $\kt{\phi_{in}}$ of a scattering system is the state of the particles a long time before the interaction, formally in the limit $t\rightarrow -\infty$.  The boundary conditions of scattering put certain constraints on the functional form of $\phi_{in}(p_1,p_2)$, and these affect the entanglement that can be generated.

First, it is assumed that the two particles are prepared independently and that before the scattering they have no correlations (classical or quantum).  Therefore, the pure in-state $\kt{\phi_{in}}$ is separable with respect to the interparticle TPS (\ref{iTPS}) and the wave function  can be factored
\begin{equation}\label{instate}
\kt{\phi_{in}} = \kt{\phi_1}\otimes\kt{\phi_2} \Leftrightarrow
\phi_{in}(p_1,p_2) = \phi_1(p_1)\phi(p_2)
\end{equation}
as a product of pure, single-particle states $\kt{\phi_i}\in\HS_i$. 
Since a global boost in momentum space is a local unitary operator, we can perform our analysis in the reference frame where the expectation value of the COM momentum $\hat{P}$ is zero.  Therefore, without any loss of generality, we will assume that  the expectation values of the single particle momentum operators in the state $\phi_{in}$ are $\langle \hat{P}_1 \rangle = k$ and $\langle \hat{P}_2 \rangle = -k$.
Further, we will also assume that the single-particle wave function $\phi_1(p_1)$ has support only (or to a good approximation) in the range $p_1>0$ and that $\phi_2(p_2)$ has support when $p_2<0$.  Then the particles are moving toward each other, and eventually transmission and reflection can be treated as modes.  This condition is implies
\begin{equation}\label{cond}
\int dp \phi_1(p) \phi_2(p) = 0.
\end{equation}
As a final assumption,the particles are initially located far from each other, or if the potential has a finite range, outside that range.
An example of such a state that satisfies these criteria is the product of Gaussian momentum wave functions (\ref{gaussmom}) with central momenta $k_1 = -k_2 = k$ a positive number, initial positions $a_2 = -a_1= a$ large and positive and where  $\sigma_i/k \ll 1$:
\begin{equation}\label{phig}
\phi_G({\bf p}) =  N_1N_2 e^{i(p_1-p_2) a}e^{-\frac{(p_1 - k)^2}{4\sigma^2_1}} e^{-\frac{(p_2 + k)^2}{4\sigma^2_2}},
\end{equation}
When $\sigma_i/k =1/6$, the integral in (\ref{cond}) is less than a millionth. 

In elastic scattering, the out-state $\kt{\phi_{out}}$, formally the state in the limit $t\rightarrow \infty$, will also be an element of $\HS_1\otimes \HS_2$.  The scattering operator $\hat{S}$ is a unitary operator on $\HS$ that connects the in-state and out-state
\begin{equation}
\kt{\phi_{out}}=\hat{S}\kt{\phi_{in}}.
\end{equation}
If the scattering potential $V_r$ depends only of the relative positions and has finite range (and also for more general assumptions), one can show that the scattering operator exists and its properties can be derived from the Schr\"odinger equation~\cite{goldberger}.  While  $\hat{S}$ preserves the purity of the total state, it is not a local operator with respect to the TPS (\ref{iTPS}).  The interaction will generate correlations between the momenta and we general expect that $\phi_{out}(p_1,p_2)\neq\phi'_1(p_1)\phi'_2(p_2)$. Then the essential problem becomes, for a given $\hat{S}$ and $\phi_{in}$, what is the interparticle purity $p_{12}(\phi_{out})$.

In one dimension, the S-matrix (the realization of the scattering operator in a certain basis) has a particularly simple form.  Using the IE CSCO, it is a sum of a transmission term and a reflection term:
\begin{eqnarray}
\br{p,q}\hat{S}\kt{p',q'}&=&S(p,p',q,q')=S_e(p,p')S_i(q,q')\nonumber\\
&=&\delta(p'-p)\left(t(q)\delta(q-q') + r(q)\delta(q+q')\right),
\end{eqnarray}
where we have factored it into an external ($p$-dependent) part and an internal ($q$-dependent) part to emphasize that it is a local operator in the IE TPS.
The functions $t(q)$ and $r(q)$ are the transmission and reflection amplitudes and by unitarity $|t(q)|^2 + |q(q)|^2=1$.  These can be determined by solving the 1-D Schr\"odinger equation in the relative momentum coordinate using the reduced mass; analytical forms exist for a variety of potentials.
Using this expression for the S-matrix, the out-state can be written as
\begin{eqnarray}
\kt{\phi_{out}}
& =& \int dp dqdp'dq' S(p,p',q,q') \check{\phi}_{in}(p',q') \kt{p,q} \nonumber\\
& =& \int dp dq \left( t(q)\check{\phi}_{in}(p,q) + r(q)\check{\phi}_{in}(p,-q)\right) \kt{p,q} \nonumber\\
&=&  \int dp dq \left( \check{\phi}_{tra}(p,q) + \check{\phi}_{ref}(p,q)\right) \kt{p,q}\nonumber\\
& =& \int dp dq \check{\phi}_{out}(p,q) \kt{p,q} 
\end{eqnarray}

To calculate the interparticle purity of the out-state, one must convert $\tilde{\phi}_{out}(p,q)$ into $\phi_{out}(p_1,p_2)$:
\begin{eqnarray}\label{eq:phout}
\kt{\phi_{out}} &= & \int dp_1 dp_2 \phi_{out}(p_1, p_2)\kt{p_1, p_2}\nonumber\\
&= &\int dp_1 dp_2 \kt{p_1, p_2}\left(\phi_{tra}(p_1, p_2) + \phi_{ref}(p_1, p_2)\right)\kt{p_1, p_2}.
\end{eqnarray}
and then insert this into (\ref{eq:mompur}).  One finds that
\begin{equation}
\phi_{tra}(p_1,p_2) = t(q_{12}) \phi_{in}(p_1,p_2)
\end{equation}
where $q_{12} = \mu_2 p_1 - \mu_1 p_2$ and
\begin{equation}
\phi_{ref}(p_1,p_2) = r(q_{12}) \phi_{in}(\overline{p}_1,\overline{p}_2)
\end{equation}
where $\overline{p}_1 = (\mu_1 - \mu_2)p_1 + 2\mu_1p_2$, and $\overline{p}_2 = 2\mu_2p_1 + (\mu_2 - \mu_1)p_2$.  The expressions for $\overline{p}_i$ arise due to the effect of transforming the particle coordinates to the COM-relative momentum coordinates, flipping the direction of the relative momentum, and then returning to the particle coordinates.  In terms of matrices, this transformation looks like
\begin{equation}
\overline{\bf p} = {\rm M} {\bf p}={\rm T}_{cm}^{-1}{\rm F}{\rm T}_{cm}{\bf p}
\end{equation}
where ${\rm F}$ is a matrix that flips the relative momentum.  In components, this is
\begin{eqnarray}\label{refl}
{\rm M} &=& \left(\begin{array}{cc}  \mu_1 & 1\\ \mu_2 & -1\end{array} \right) \left(\begin{array}{cc} 1 & 0\\ 0 & -1\end{array} \right) \left(\begin{array}{cc} 1 & 1\\ \mu_2 & -\mu_1\end{array} \right) \nonumber\\
 &=& \left(\begin{array}{cc}  (\mu_1 - \mu_2) & 2\mu_1\\ 2\mu_2 & (\mu_2 - \mu_1)\end{array} \right).
\end{eqnarray}

One can prove that the transformation (\ref{refl}) maps disjoint regions of the $(p_1,p_2)$-plane into disjoint regions.  In particular, it reflects the line $p_1 = -p_2$ about the origin without change of scale, i.e. $(k,-k)\rightarrow(-k,k)$.  This means that $\phi_{tra}(p_1,p_2)$ and $\phi_{ref}(p_1,p_2)$ have disjoint domains of support; i.e. they are `modes' in the sense of the introduction. When calculating the integral in $p_{12}(\phi_{out})$, one takes the four-fold product of $\phi_{out}$ with different permutations of $p_1$, $p'_1$, $p_2$ and $p'_2$.  The only terms that survive the integrals over all four variables are those terms which involve only $\phi_{tra}$ or only $\phi_{ref}$.  In other words, the integral for $p_{12}(\phi_{out})$ breaks into two parts, the purity of the transmitted state and the purity of the reflected state:
\begin{equation}\label{split}
p_{12}(\phi_{out}) = p_{12}(\phi_{tra})+p_{12}(\phi_{ref}).
\end{equation}
Note that since the probability for transmission $T=||\phi_{tra}||^2$ and reflection $R=||\phi_{ref}||^2$ are less then one, one can show $p_{12}(\phi_{tra})\in(0,T^2]$ and $p_{12}(\phi_{ref})\in(0,R^2]$, so $p_{12}(\phi_{out})$ is bounded from above by one.

\section{Useful Approximations}

Very few assumptions went into (\ref{split}): the interaction was Galilean invariant and sufficiently well-behaved that the S-operator exists, and the in-state wave function must satisfy certain boundary conditions typical of scattering experiments.  If $\phi_{in}$ and $\hat{S}$ are explicitly known, then (\ref{split}) can be calculated numerically.  However, there are further useful analytic approximations that are applicable under reasonable circumstances.

\subsection{Entanglement of reflection}

The first approximation applies when the transition amplitudes $t(q)$ and $r(q)$ do not vary much with $q$.  Note that since we calculate in the COM frame, the transformation (\ref{spmtoiem}) implies the expectation value of $q$ is $k$ and the uncertainty of the relative momentum $q$ is $\Delta q = \sqrt{(\mu_1 \Delta p_1)^2 + (\mu_2 \Delta p_2)^2} $.  If we assume that $t(k)\gg\Delta q \partial t/\partial q $, then we can approximate the functions $t(q_{12})$ and $r(q_{12})$ by their values at the expectation value of relative momentum, $t(k)$ and $r(k)$.
With this assumption, the purity of the transmitted state can be calculated immediately:
\begin{eqnarray}\label{traen}
p_{12}(\phi_{tra}) &= &\int dp_1dp_2dp'_1dp'_2 \phi_{tra}(p_1, p_2)\phi_{tra}^*(p'_1, p_2)\phi_{tra}(p'_1, p'_2)\phi_{tra}^*(p_1, p'_2)\nonumber\\
&= & |t(k)|^4\int dp_1dp_2dp'_1dp'_2 \phi_{in}(p_1, p_2)\phi_{in}^*(p'_1, p_2)\phi_{in}(p'_1, p'_2)\phi_{in}^*(p_1, p'_2)\nonumber\\
&= & |t(k)|^4 p_{12}(\phi_{in}) = T^2
\end{eqnarray}
because $p_{12}(\phi_{in})=1$.  If $t(q)$ does vary rapidly, then the integral cannot be simplified because the factors like $t(q_{12})$ will generally lead to a marked distortion of the wave function that does not preserve separability.  This is particularly true when there is a resonance and $t(q)$ varies to a maximum value of 1 over a small range of $q$~\cite{tal05}.

The purity of the reflected state cannot be simplified in such a way even if $r(q)$ is a constant. The inversion of the internal momentum $q$  leads to entanglement between $p_1$ and $p_2$.  The reflected wave function is
\begin{equation}
\phi_{ref}(p_1,p_2) = r(q_{12})\phi_{in}(({\rm M}\vp)_1, ({\rm M}\vp)_2) = r(q_{12})\phi_{\overline{in}}(p_1,p_2),
\end{equation}
where ${\rm M}$ is the matrix (\ref{refl}) that enacts the transformation on $(p_1,p_2)$ due to $q\rightarrow -q$.
One can think of this as another example of a linear coordinate transformation described in Section II, except now this is an active transformation of the wave function instead of a passive transformation of the coordinates.  One can use the formula (\ref{eq:zpur1}) to calculate the covariance properties interparticle purity for a state $\phi_{\overline{in}}(p_1,p_2)$. By making the the substitution $d\overline{p}_1d\overline{p}_2=dp_1 dp_2$, one finds
\begin{equation}
p_{12}(\phi_{\overline{in}})=p_{{\rm M}12}(\phi_{in}).
\end{equation}

Assuming a Gaussian in-state of the form (\ref{phig}), one then finds
\begin{equation}\label{barin}
p_{12}(\phi_{\overline{in}})=\frac{\sigma_1 \sigma_2}{\sqrt{((\mu_1 - \mu_2)^2\sigma_1^2 + 4\mu_1^2\sigma_2^2)(4\mu_2^2\sigma_1^2 + (\mu_2 - \mu_1)^2\sigma_2^2)}}.
\end{equation}
This is a function of the masses and momentum uncertainties of the in-state single-particle wave functions, but it can also be re-expressed entirely in terms of the ratios $m_2/m_1$ and $\sigma_2/\sigma_1$.  The scale of the entanglement by reflection is set by the relative masses and momentum uncertainties of the scattering particles.
The purity takes its maximum value (and therefore there is the least entanglement) when the masses are either equal $m_1 = m_2$ or when the Schulman  condition $m_1/\sigma_1^2 = m_2/\sigma_2^2$ is satisfied.  
In \cite{janzig06}, the authors identify these same two conditions as necessary for non-entangling scattering from hard core potentials.  Since hard core scattering assumes only reflection for all $q$, we see that any entanglement generated by hard core scattering is of a purely kinematic origin and independent of the energy scale of the dynamics.  Reflection is an entangling process except for equal mass scattering or for when the Schulman condition is satisfied.  This effect does not disappear even in the limit of narrow momentum distributions.

For all other potentials, there can be both transmission and reflection, and therefore both kinematic and dynamic entanglement even in the constant amplitude approximation.  The total interparticle entanglement in this approximation can be efficiently calculated using the sum
\begin{equation}
p_{12}(\phi_{out}) = T^2 + R^2 p_{12}(\phi_{\overline{in}}).
\end{equation}
with $T+R =1$ and $ p_{12}(\phi_{\overline{in}})$ given by (\ref{barin}).  Note that the strength of the dynamical entanglement depends on the transition probabilities, which in turn depend on the energy scale of set by the scattering particles and the interaction, unlike the kinematic entanglement of reflection which is independent of the overall energy scale.

\subsection{Coarsest approximation: Two two-level systems}

In any experiment, there is incomplete measurement of the out-state and one does not have access to full information about the wave function.  Real experiments cannot measure probability distributions over continuous variables; the continuous variable must be broken into `bins' with some resolution.  This coarse-graining is equivalent to discretizing the system.  In the simplest model where the continuous variable of interest is broken into $N$ disjoint bins, then each one-particle system may be approximated as an $N$-dimensional discrete system.  Then the minimum purity possible for a coarse-grained bipartite continuous-variable system would be $1/N$, whereas the `unbinned' purity is limited from below by zero.  

For example, in the worst case, it might only be possible to detect whether after scattering the distinguishable particles remain on their original sides or they have transmitted past each other.  Then one cannot measure the all correlations in the two-particle wave function $\phi_{out}(p_1,p_2)$ that may exist, but one only knows that the function has support around one of two domains: the transmitted mode centered around $(k,-k)$ and the reflected mode centered around $(-k,k)$.  
Then the particles are effectively two level systems.  In this approximation, each particle has two states that we label as $\kt{\pm}$, the sign of its momentum in the COM frame.  For the in-state, we have
\begin{equation}
\kt{\phi_{in}}=\kt{+}\otimes\kt{-}=\kt{+-}.
\end{equation}
The out-state will in general be a superposition of the transmitted and reflected states
\begin{equation}
\kt{\phi_{out}}=\tilde{t} \kt{+-}+\tilde{r} \kt{-+}.
\end{equation}
The numbers $\tilde{t}$ and $\tilde{r}$ are the effective amplitudes chosen such that $|\tilde{t}|^2=T$ and $|\tilde{r}|^2=R$.  If the transmission and reflection amplitudes are relatively constant over $q$ around $k$, then $\tilde{t}\approx t(k)$ and $\tilde{r}\approx r(k)$, but this approximation works in the more general case as long as the $\phi_{ref}(p_1,p_2)$ and $\phi_{tra}(p_1,p_2)$ are essentially orthogonal.

In this approximation, the space of states for each particle is effectively realized by $\mathbb{C}^2$ and the total Hilbert space $\HS$ for the system is realized as $\HS=\HS_1\otimes\HS_2\sim\mathbb{C}^2\times\mathbb{C}^2=\mathbb{C}^4$.  We can calculate the interparticle entanglement by looking at the reduced density matrix for one of the particles:
\begin{equation}
\rho_1 = \Tr_2(\kb{\phi}{\phi}).
\end{equation}
For the in-state, we have
\begin{equation}
\rho_1 = \left(\begin{array}{cc} 1&0\\0&0\end{array}\right)
\end{equation}
and then we can calculate the entanglement by the purity of the reduced density matrix $p_{12}=\Tr_1(\rho_1^2) = 1$.  As expected there is no entanglement.  For the out-state, we have
\begin{equation}
\rho_1 = \left(\begin{array}{cc} T&0\\0&R\end{array}\right).
\end{equation}
The entanglement of the particles in this ``no wave function'' approximation is
\begin{equation}
p_{12} = T^2 + R^2.
\end{equation}
Interparticle entanglement is maximized when $T = R = 1/2$. Then $p_{12}= 1/2$, which is the same maximum entanglement as for two two-level systems.  Physically, the superposition of transmission and reflection leads to the entanglement of the particles, and when these components have equal likelihood, the particles are most entangled by the collision.  Not just correlation is required for interparticle entanglement (the in-state had perfect anti-correlation between the particle velocities), but indeterminacy between the two modes of transmission and reflection is required if one cannot measure the correlations between the particles within a mode.  

\section{Conclusion}

Our primary concern has been the interparticle entanglement of the pure out-state.  However, there are several factors that limit how measurable this entanglement could be.  First, typically the in-state will be a mixed state $\rho_{in}$.  The boundary conditions on the in-state still require it to be separable, but purity will no longer be an acceptable measure of the entanglement generated in the out-state $\hat{S}\rho_{out}\hat{S}^\dag$.  Second, scattering experiments extract information about the out-state from transition rates and cross sections.  The measuring devices can be represented as operators $\Lambda$, and then measurements are connected to the probabilities by the Born rule $\mathrm{prob} = \mathrm{Tr}(\rho_{out}\lambda)$.  Since there are no perfect measuring devices, the kind of coarse-graining described above will always reduce the measurable entanglement from its theoretical maximum.  Finally, as extensive research on Bell inequalities has shown, some of the correlations measured in a scattering experiment are classical, or they may be quantum but do not rise above some threshold that could not be explained by a local, realistic theory.  See \cite{mack02,frey07} for further discussion of this last point.

Aside from issues of direct measurement of entanglement due to scattering, another interesting avenue of research is the connection between entanglement in individual scattering events and entanglement properties of systems of particles.  For example, scattering dynamics give a contribution to the equation of state for effective many-body Hamiltonians of systems like ultracold atomic gases.  Such analyses usually assume that the exact form of the interaction is unimportant, but as seen above, kinematic effects may still drive the generation of entanglement in scattering.  For example,  L.S.~Schulman discusses in \cite{schulman04} the case of two gases of atoms with different masses interacting through scattering.  As a result of the kinematic entanglement due to reflection, he derives the result that the interparticle entanglement for any interspecies pair reduces to zero after multiple scatterings and that the particle wave functions tend toward Gaussian distributions that satisfy the Schulman condition (\ref{sec}).  How to implement this hypothesis experimentally is unclear, but advances in the control of ultracold atomic systems promise future novel experiments in interparticle entanglement.

\begin{acknowledgments}
The authors would like to thank the Research Corporation for supporting this research through a Cottrell College Science Award and to thank Amit Kapadia for fruitful discussions.
\end{acknowledgments}

\section*{Appendix}

Here we present a derivation of (\ref{pur_trans1}).  Our method is similar to that found in \cite{janzig06} which is inspired by squeezed-state quantum optics calculations, but we use Gaussian wave functions, not Gaussian Wigner functions.
A useful result will be that for an $N\times N$ diagonal matrix ${\rm B}$ with elements $B_{ii}$ and an $N$-dimensional real vector ${\bf x}$, we have the Gaussian integral 
\begin{equation}\label{general}
\int_{-\infty}^{\infty} d^N{\bf x} \exp(-{\bf x}^T {\rm B} {\bf x}) = \prod_{i=1}^N \sqrt{\frac{\pi}{B_{ii}}}.
\end{equation}

First we rewrite (\ref{gaussmom}) as 
\begin{equation}\label{phigg}
\phi({\bf p}) = N_1 N_2 \exp(i{\bf a} \cdot\vp  )\exp(-\frac{1}{2}(\vp-\vk)^T {\rm V} (\vp-\vk))
\end{equation}
where $N_i = (2\pi \sigma_1^2)^{-1/4}$, ${\bf a} = ( a_1 ,a_2)$, $\vk = (k_1, k_2)$ and ${\rm V}$ is the $2\times 2$ matrix
\begin{equation}
{\rm V} = \left( \begin{array}{cc} \frac{1}{2\sigma^2_1} & 0 \\ 0 & \frac{1}{2\sigma^2_2}\end{array}\right).
\end{equation}

The new coordinates are defined by ${\rm T}{\bf p}={\bf z}$.
These coordinates imply a different CSCO $\{\hat {Z}_1, \hat{Z}_2\}$ and therewith a new TPS.  We want to calculate the entanglement of (\ref{phigg}) in this new TPS.  Using the notation $\tilde{\rm T}= {\rm T}^{-1}$, the wave function in the new CSCO is
\begin{eqnarray}\label{phiggz}
\check{\phi({\bf z})} &= &N_1 N_2 \exp(i {\bf a}\cdot\tilde{\rm T} \vz )\exp(-\frac{1}{2}(\tilde{\rm T}\vz-\vk)^T {\rm V} (\tilde{\rm T}\vz-\vk))\nonumber\\
 &= &N_1 N_2 \exp(i {\bf a}\cdot\tilde{\rm T} \vz )\exp(-\frac{1}{2}\vz^T \tilde{\rm V} \vz)\exp(-\vk{\rm V}\tilde{\rm T}\vz)\exp(-\frac{1}{2}\vk^T {\rm V}\vk),
\end{eqnarray}
where $\tilde{\rm V} = \tilde{\rm T}^T V \tilde{\rm T}$.

As a next step, we define the following variables: ${\bf z}=(z_1,z_2)$, ${\bf z}'=(z'_1,z'_2)$, $\overline{\bf z}=(z'_1,z_2)$, $\overline{\bf z}'=(z_1,z'_2)$.  Then, the purity $P_{{\rm T}12}$ is
\begin{eqnarray}
p_{\mathrm{T}12}(\phi_G) &= &\int d^2{\rm z}d^2{\rm z}'  \check{\phi({\bf z})}\check{\phi(\overline{\bf z})}^*\check{\phi({\bf z}')}\check{\phi(\overline{\bf z}')}^*\nonumber\\
&= & N_1^4N_2^4 \int d^2{\rm z}d^2{\rm z}'  \exp(i {\bf a}\cdot\tilde{\rm T} (\vz - \overline{\bf z} + \vz' -\overline{\vz}') \exp(-2\vk^T {\rm V}\vk)\nonumber\\
&&\times \exp(-\frac{1}{2}(\vz^T \tilde{\rm V} \vz + \overline{\vz}^T \tilde{\rm V} \overline{\vz}) + (\vz')^T \tilde{\rm V} \vz' + (\overline{\vz}')^T \tilde{\rm V} \overline{\vz}')\exp(-\vk{\rm V}\tilde{\rm T}(\vz + \overline{\bf z} + \vz' +\overline{\vz}')\nonumber\\
&= & N_1^4N_2^4 \int d^2{\rm z}d^2{\rm z}'  \exp(-2\vk^T {\rm V}\vk-2\vk{\rm V}\tilde{\rm T}(\vz + \vz')\nonumber\\
&&\times  \exp(-\frac{1}{2}(\vz^T \tilde{\rm V} \vz + \overline{\vz}^T \tilde{\rm V} \overline{\vz} + (\vz')^T \tilde{\rm V} \vz' + (\overline{\vz}')^T \tilde{\rm V} \overline{\vz}').\label{mid}
\end{eqnarray}

To integrate this in the fashion of (\ref{general}), we define the following variables: ${\rm w} = \tilde{\rm T} - \vk$, ${\rm y} = \vz - \vz'$, and
\begin{equation}
{\rm D} = \left( \begin{array}{cc}  \tilde{\rm V}_{11} & 0 \\ 0 &  \tilde{\rm V}_{22}\end{array}\right)
\end{equation}
where $\tilde{\rm V}_{11} = u^2 V_{11} + t^2 V_{22}$ and  $\tilde{\rm V}_{22} = s^2 V_{11} + r^2 V_{22}$.
By defining ${\rm D}$ as just the diagonal elements of $\tilde{\rm V}$, the exponential terms in (\ref{mid}) can be re-expressed as two purely quadratic terms:
\begin{equation}
p_{\mathrm{T}12}(\phi_G) =  \frac{1}{4} N_1^4N_2^4 \int d^2{\rm w}d^2{\rm y} \exp(-\frac{1}{2}({\bf w}^T {\rm V} {\bf w}))\exp(-\frac{1}{2}({\bf y}^T {\rm D} {\bf y})) ,
\end{equation}
where the fact that $4 d^2{\rm z}d^2{\rm z}'=\int d^2{\rm w}d^2{\rm y}$ has been used.  Finally, using (\ref{general}), $p_{\mathrm{T}12}$ can be evaluated
\begin{eqnarray}
p_{\mathrm{T}12}(\phi_G) &= &\frac{1}{4} N_1^4N_2^4 \sqrt{\frac{2\pi}{V_{11}}}\sqrt{\frac{2\pi}{V_{22}}}\sqrt{\frac{2\pi}{D_{11}}} \sqrt{\frac{2\pi}{D_{22}}}\nonumber\\
&=& \frac{\sigma_1 \sigma_2}{\sqrt{(r^2\sigma_1^2 + s^2\sigma_2^2)(t^2\sigma_1^2 + u^2\sigma_2^2)}}.
\end{eqnarray}

\end{document}